\documentclass[twocolumn,showpacs,preprintnumbers,prl,superscriptaddress]{revtex4}
\usepackage{graphicx}
\usepackage{amssymb}
\usepackage{amsmath}

\begin{document}

\title{Electron self-trapping and fluctuation density-of-states tail at the
critical point}
\author{M. I. Auslender}
\affiliation{Ben-Gurion University of the Negev, POB 653, Beer Sheva 84105, Israel}
\author{M. I. Katsnelson}
\affiliation{Institute for Molecules and Materials, Radboud University of Nijmegen,
NL-6525 ED Nijmegen, The Netherlands}
\date{\today}

\begin{abstract}
We consider electron self-trapping due to its interaction with
order-parameter fluctuations at the second-order phase-transition or
critical point (for example, at the Curie temperature in magnetic or
ferroelectric semiconductors). Using Feynman path integral approach the
autolocalization energy and the size of the self-trapped state (fluctuon)
are estimated. It is shown that the fluctuon states are connected with the
Lifshitz tail of the electron density-of-states, the parameters of this tail
being determined by the critical exponents.
\end{abstract}

\maketitle

The interaction of the charge carrier in a semiconductor with some
order-parameter fluctuations can drastically change its state
leading to a self-trapping, or autolocalization
\cite{brinkman,krivoglaz,ourTMF1,ourJMMM,ourTMF2,nagaev,dagotto}.
This phenomenon is of crucial importance, for example, for so hot
subject as phase separation in magnetic semiconductors and
colossal magnetoresistance materials
\cite{ourTMF2,nagaev,dagotto}, where the magnetization plays the
role of the order parameter. Since the band motion of the electron
is easier (and, hence, the bandwidth is larger) for
ferromagnetically ordered state the electron in antiferromagnetic
or magnetically disordered surrounding creates a ferromagnetic
region (magnetic polaron, \textquotedblleft
ferron\textquotedblright\ \cite{nagaev}, or \textquotedblleft
fluctuon\textquotedblright\ \cite{krivoglaz}) and turns our to be
self-trapped in this region. Recently, the formation of the
magnetic polaron in ferromagnetic semiconductors EuB$_6$ and,
possibly, EuO, has been observed (see Refs.\onlinecite{Eu1,Eu2}
and references therein). The order parameter can be also of
different origin, e.g., electric polarization in ferroelectric
semiconductors, or crystallographic order parameter in ordered
alloys \cite{krivoglaz}. Phase transitions in helium monolayers
\cite{He} and ultracold atom systems in optical lattices
\cite{BEC} might be novel interesting examples.

The \textquotedblleft driving force\textquotedblright\ of the
self-trapping is always a band narrowing in disordered state in
comparison with the completely ordered one. This intuitive view
\cite{mott} can be confirmed by a rigorous consideration for the
classical $s-d$ exchange (\textquotedblleft Kondo
lattice\textquotedblright ) model on the Bethe lattice
\cite{ourTMF2}; however, for the real lattices it is actually not
very accurate. It was demonstrated already in a seminal paper by
Brinkman and Rice \cite{brinkman} that the band \textit{edges} do
not depend on the degree of spin disorder and the band narrowing
means rather the decrease of the electron density-of-states (DOS)
momenta, the DOS near the edges being exponentially small in the
disordered state (Lifshitz tails \cite{lifshitz,VK}). The
fluctuons (we will use this term to emphasize that the magnetic
origin of the order parameter is actually not relevant for the
problem under consideration) are connected with the Lifshitz
tails.

The autolocalization region becomes larger when the energy goes
closer to the band edge (and the edge itself corresponds to the
state with complete ordering in the whole crystal)
\cite{brinkman}. The consideration of the fluctuon problem is
crucially dependent on the ratio of the autolocalization
radius $l$ to another scale of the system, namely, the correlation length $%
\xi $. In particular, a standard phenomenological consideration \cite%
{krivoglaz} is inapplicable at the critical point (or the point of the
second-order phase transition), where $\xi \rightarrow \infty $. At the same
time, this case is especially interesting since the fluctuations near the
critical point are strongest and the self-trapping conditions fulfill much
easier in this case. Here we present a theory for the Lifshitz DOS tails and
fluctuon states at the critical point $T=T_{c}$.

Following Refs. \onlinecite{ourTMF1,ourJMMM} we will use the path
integral variational approach developed by Feynman for the problem
of polaron in ionic crystal \cite{feynman,path1}. For simplicity,
we start with the case of a scalar order-parameter acting only on
the orbital motion of the electron and not on its spin (for
example it may be the critical point in ferroelectrics); some
generalization will be considered further. Then, in continuum
approximation, the Hamiltonian of the system consisting of the
electron and the order-parameter field can be written in a simple
form
\begin{equation}
\mathcal{H}=\mathcal{H}_{f}\left( \varphi \right) +\mathcal{H}_{e}\left(
\mathbf{r,}\varphi \right) ,\;\mathcal{H}_{e}\left( \mathbf{r,}\varphi
\right) =-\frac{1}{2}\nabla _{\mathbf{r}}^{2}-g\varphi \left( \mathbf{r}%
\right)  \label{ham}
\end{equation}%
where we have chosen the units $\hbar =m=1$, $m$ is the electron effective
mass, $\mathbf{r}$ is the electron coordinate, $\varphi \left( \mathbf{r}%
\right) $ is the order-parameter field with its own Hamiltonian $\mathcal{H}%
_{f}\left( \varphi \right) $ and $g$ is the coupling constant; we
will choose it in such a way that $\varphi $ varies between $-1$
and $1$. The partition function of the whole system may be
transformed to
\begin{align}
Z=& \mbox{Tr}e^{-\beta \mathcal{H}_{f}\left( \varphi \right) -\beta \mathcal{%
H}_{e}\left( \mathbf{r,}\varphi \right) }  \notag \\
=& Z_{f}\left\langle \mbox{Tr}_{\mathbf{r}}T_{\tau }\exp \left[
-\int_{0}^{\beta }\mathcal{H}_{e}\left( \mathbf{r,}\varphi \left( \mathbf{r,}%
\tau \right) \right) d\tau \right] \right\rangle _{f}  \label{partfun}
\end{align}%
where $\beta =T^{-1}$ is the inverse temperature, $Z_{f}=\mbox{Tr}_{\varphi
}e^{-\beta \mathcal{H}_{f}\left( \varphi \right) }$ is the partition
function of the field, $\varphi \left( \mathbf{r,}\tau \right) =e^{\tau
\mathcal{H}_{f}\left( \varphi \right) }\varphi \left( \mathbf{r}\right)
e^{-\tau \mathcal{H}_{f}\left( \varphi \right) }$ and
\begin{equation}
\left\langle \mathcal{A}\left( \varphi \right) \right\rangle _{f}=\frac{1}{%
Z_{f}}\mbox{Tr}_{\varphi }e^{-\beta \mathcal{H}_{f}\left( \varphi \right) }%
\mathcal{A}\left( \varphi \right)  \label{meanf}
\end{equation}%
is the average over the field states. Here we will consider only classical
case where $\tau $-dependence of the field can be neglected. Using Feynman
path-integral approach \cite{path1,path2,path3} and taking average over $%
\varphi $ yields for the electron-only free energy
\begin{equation}
\mathcal{F}=-\frac{1}{\beta }\left( \ln Z-\ln Z_{f}\right) =-\frac{1}{\beta }%
\ln \int\limits_{\mathbf{r}\left( 0\right) =\mathbf{r}\left( \beta \right)
}e^{-\mathcal{S}}\mathcal{D}\left[ \mathbf{r}\left( \tau \right) \right] ,
\label{freen}
\end{equation}%
where $\mathcal{S}_{0}+\mathcal{S}_{int}$ is the effective action,
\begin{eqnarray}
\mathcal{S}_{0} &=&\frac{1}{2}\int_{0}^{\beta }\left[ \overset{\bullet }{%
\mathbf{r}}\left( \tau \right) \right] ^{2}d\tau  \label{S} \\
\mathcal{S}_{int} &=&-\sum_{m=2}^{\infty }\frac{g^{m}}{m!}\int_{0}^{\beta
}...\int_{0}^{\beta }\mathcal{K}_{m}\left( \mathbf{r}\left( \tau _{1}\right)
...\mathbf{r}\left( \tau _{m}\right) \right) d\tau _{1}...d\tau _{m}  \notag
\end{eqnarray}%
and $\mathcal{K}_{m}\left( \mathbf{r}_{1};...;\mathbf{r}_{m}\right) $ are
the $m$-th cumulant correlators, defined recursively by
\begin{align}
\mathcal{K}_{1}\left( \mathbf{r}_{1}\right) & =\left\langle \varphi \left(
\mathbf{r}_{1}\right) \right\rangle _{f},  \notag \\
\mathcal{K}_{2}\left( \mathbf{r}_{1};\mathbf{r}_{2}\right) & =\left\langle
\varphi \left( \mathbf{r}_{1}\right) \varphi \left( \mathbf{r}_{2}\right)
\right\rangle _{f}-\mathcal{K}_{1}\left( \mathbf{r}_{1}\right) \mathcal{K}%
_{1}\left( \mathbf{r}_{2}\right) ,...  \label{cumulant}
\end{align}%
etc. Further we consider only the case of $\mathcal{K}_{1}=0$.

To estimate $\mathcal{F}$ we use the same trial action as in Refs.\onlinecite%
{ourTMF1,ourJMMM}, $\mathcal{S}_{t}=\mathcal{S}_{0}+\mathcal{S}_{pot}$ where
\begin{equation}
\mathcal{S}_{pot}=\frac{\omega ^{2}}{4\beta }\int_{0}^{\beta
}\int_{0}^{\beta }\left[ \mathbf{r}\left( \tau \right) -\mathbf{r}\left(
\sigma \right) \right] ^{2}d\tau d\sigma ,  \label{tract}
\end{equation}%
the oscillator frequency $\omega $ being trial parameter (in
contrast with Ref.\onlinecite{feynman} we do not introduce any
retardation in the trial action since our field is supposed to be
static). Then the Peierls-Feynman-Bogoliubov inequality applied to
Eq. (\ref{freen}) reads
\begin{equation}
\mathcal{F}\leq \mathcal{F}_{t}+\frac{1}{\beta }\left\langle \mathcal{S}%
_{int}-\mathcal{S}_{pot}\right\rangle _{t}  \label{PBineq}
\end{equation}%
where
\begin{align}
\mathcal{F}_{t}& =-\frac{1}{\beta }\ln \int_{\mathbf{r}\left( 0\right) =%
\mathbf{r}\left( \beta \right) }e^{-\mathcal{S}_{t}}\mathcal{D}\left[
\mathbf{r}\left( \tau \right) \right] ,\;  \notag \\
\left\langle \mathcal{A}\right\rangle _{t}& =\int_{\mathbf{r}\left( 0\right)
=\mathbf{r}\left( \beta \right) }\mathcal{A}\left[ \mathbf{r}\left( \tau
\right) \right] e^{\beta \mathcal{F}_{t}-\mathcal{S}_{t}}\mathcal{D}\left[
\mathbf{r}\left( \tau \right) \right] ,  \label{trfun}
\end{align}%
which is equivalent to
\begin{align}
\mathcal{F}& \leq \mathcal{F}_{t}-\frac{\omega ^{2}}{4\beta ^{2}}%
\int_{0}^{\beta }\int_{0}^{\beta }\left\langle \left[ \mathbf{r}\left( \tau
\right) -\mathbf{r}\left( \sigma \right) \right] ^{2}\right\rangle _{t}d\tau
d\sigma  \notag \\
& -\sum_{m=2}^{\infty }\frac{g^{m}}{m!\beta }\int_{0}^{\beta
}...\int_{0}^{\beta }\left\langle \mathcal{K}_{m}\left( \mathbf{r}\left(
\tau _{1}\right) ,...,\mathbf{r}\left( \tau _{m}\right) \right)
\right\rangle _{t}\prod_{j=1}^{m}d\tau _{j}  \label{interim1}
\end{align}

To proceed, we will pass to the Fourier transforms of the cumulants $%
\mathcal{K}_{m}\left( \mathbf{K}_{1},..,\mathbf{K}_{m-1}\right) $ and take
into account that for the Gaussian trial action $\mathcal{S}_{t}$ one has
\begin{align}
& \left\langle \exp \left\{ i\sum_{j=1}^{m-1}\mathbf{K}_{j}\left[ \mathbf{r}%
\left( \tau _{j}\right) -\mathbf{r}\left( \tau _{m}\right) \right] \right\}
\right\rangle _{t}=  \notag \\
& \exp \left\{ -\sum_{j,k=1}^{m-1}\frac{\mathbf{K}_{j}\mathbf{K}_{k}}{2D}%
\left\langle \left[ \mathbf{r}\left( \tau _{j}\right) -\mathbf{r}\left( \tau
_{m}\right) \right] \left[ \mathbf{r}\left( \tau _{k}\right) -\mathbf{r}%
\left( \tau _{m}\right) \right] \right\rangle _{t}\right\} ,
\label{interim3}
\end{align}%
where $D$ is the space dimensionality, $\mathbf{K}_{j}$ are the wave-vectors.

For the states in the tail the variational parameter $\omega $ satisfies the
inequalities $\beta \omega \gg 1$ and $\omega <<W=\frac{1}{2}K_{\max }^{2}$
where $K_{\max }$ is the Debye wave vector and $W$ is of order of the
electron bandwidth (the last inequality just means that the fluctuon size $%
l=1/\sqrt{2\omega }$ is much larger than interatomic distance). Thus we have
\cite{ourTMF1,ourJMMM}
\begin{align}
& \mathcal{F}\leq \frac{D}{4}\omega -\sum\limits_{m=2}^{\infty }\frac{1}{m!}%
\left( g\beta \right) ^{m}\int ...\int \mathcal{K}_{m}\left( {\mathbf{K}}%
_{1},...,{\mathbf{K}}_{m-1}\right) \times   \notag \\
& \exp \left[ {-\frac{1}{{4}\omega }\sum\limits_{j=1}^{m}{\mathbf{K}_{j}^{2}}%
-\frac{1}{{4}\omega }\left( {\sum\limits_{j=1}^{m}{\mathbf{K}_{j}}}\right)
^{2}}\right] \prod\limits_{j=1}^{m-1}{\frac{{\Omega _{D}d^{D}K_{j}}}{{\left(
{2\pi }\right) ^{D}}}}
\end{align}%
where ${\Omega _{D}}$ is the unit lattice volume. If one can neglect the ${%
\mathbf{K}}$-dependence of the cumulants this sum can be transformed \cite%
{ourTMF1,ourJMMM} to the phenomenological expression \cite{krivoglaz} in
terms of the fluctuation free energy in \textit{homogeneous} field created
by the electron. This assumption works not too close to the critical point
where $l\gg \xi $. At the critical point this assumption can be never used.
Instead, the scaling properties hold \cite{PP}
\begin{equation}
\mathcal{K}_{m}\left( {\mathbf{K}}_{1},...,{\mathbf{K}}_{m-1}\right)
=a^{\left( 2-\eta \right) \left( m-1\right) }\mathcal{K}_{m}\left( a{\mathbf{%
K}}_{1},...,a{\mathbf{K}}_{m-1}\right)
\end{equation}%
where $\eta $ is the anomalous dimensionality critical exponent. Making
replacement of variables by $\mathbf{K}_{j}=\sqrt{q}\mathbf{\varkappa }_{j}$%
, where $q=\omega /W$, and further returning to the real space coordinates
conjugate to $\mathbf{\varkappa }_{j}$, we get
\begin{align}
\mathcal{F}& \leqslant \frac{DW}{4}q-q^{-d/2}\beta
^{-1}\sum\limits_{m=2}^{\infty }{\frac{\left( \beta q^{d/2}\right) {^{m}}}{{%
m!}}}\times   \notag \\
& \int {...\int {K_{m}}\left( {{\mathbf{r}}_{1},...,{\mathbf{r}}_{m}}\right)
}\prod\limits_{j=1}^{m}u_{D}{\left( r{_{j}}\right) }d^{D}r_{j}  \notag \\
& =\frac{DW}{4}q-q^{-d/2}f\left[ q^{d/2}u_{D}\left( r\right) \right] \hfill ,
\label{sum}
\end{align}%
where $d=D-2+\eta $ is the anomalous space dimensionality,
\begin{equation}
u_{D}\left( r\right) =g\left( \frac{K_{\max }}{2\pi }\right) ^{D/2}\exp
\left( {-}\frac{1}{2}K_{\max }^{2}r^{2}\right)   \label{h}
\end{equation}%
is the potential localized within Debye sphere around $\mathbf{r}=0$;  $f[U(%
\mathbf{r})]$ is the change of the free energy of the fluctuations bath upon
switching an external potential $U(\mathbf{r})$ on.

Earlier \cite{ourTMF1} we considered the Gaussian approximation (only the
term with $m=2$ in Eq.(\ref{sum})) which is applicable for not too large
coupling constants. To consider the states near the band edge $E=-g$ we have
to sum up the series (\ref{sum}). To this end, we employ the fact that $%
u_{D}\left( r\right) $ is virtually $D$-dimensional delta function for
typical $r\gtrsim l$. Further, to avoid ultraviolet divergencies, we return
to the discrete-lattice Ising model, where $\varphi _{\mathbf{r}_{i}}=\pm 1$%
. In this framework we adopt that the above potential acts only at site $%
\mathbf{r}_{i}=\mathbf{0}$, viz. $u_{D}\left(
\mathbf{r}_{i}\right) =g \delta _{\mathbf{r}_{i}\mathbf{,0}}$.
This assumption can only affect some numerical coefficients of
order of unity in the following estimations. Then in the limit of
$gq^{d/2}\gg 1$ corresponding to the states near the band edge one
obtains
\begin{equation}
f\left[ q^{d/2}u_{D}\left( r\right) \right] \simeq -gq^{d/2}+\frac{1}{\beta }%
\ln G_{D}  \label{f}
\end{equation}%
where $G_{D}=\left\langle \delta _{\varphi \left( \mathbf{0}\right)
,1}\right\rangle _{f}^{-1}.$ This is just a number larger than $1$ which can
be calculated if all necessary correlation functions are known. For the
Ising model with nearest-neighbor interaction $J$, one can easily derive
\begin{equation}
G_{D}^{-1}=\exp \left( \beta _{c}\mathcal{F}_{f}\right) \left\langle
\prod\limits_{i=1}^{z}\left( \cosh \beta _{c}J+\varphi _{\mathbf{r}%
_{i}}\sinh \beta _{c}J\right) \right\rangle _{f}  \label{G_D-Ising}
\end{equation}%
where $z$ is the nearest-neighbor number and $\mathcal{F}_{f}$ is the Ising
model free energy per site at $T=T_{c}=\beta _{c}^{-1}$.

Substituting Eq.(\ref{f}) into Eq.(\ref{sum}) and minimizing with respect to
$q$ one finds for the optimal fluctuon size
\begin{equation}
l_{0}=\frac{1}{K_{\max }\sqrt{q_{0}}}=\frac{1}{K_{\max }}\left( \frac{W}{%
T_{c}}\frac{D}{2d\ln G_{D}}\right) ^{\frac{1}{d+2}}  \label{size}
\end{equation}%
and for the fluctuon energy
\begin{equation}
\mathcal{F}_{0}=-g+\frac{d+2}{2}\left( T_{c}\ln G_{D}\right) ^{\frac{2}{d+2}%
}\left( \frac{DW}{2d}\right) ^{\frac{d}{d+2}}.  \label{energy}
\end{equation}%
The fluctuon forms only at $\mathcal{F}_{0}<0$. This requirement yields
necessary condition for the self-trapping, which reads%
\begin{equation}
\frac{T_{c}}{W}<\frac{D}{2d\ln G_{D}}\left( \frac{2}{D+\eta }\frac{2d}{D}%
\frac{g}{W}\right) ^{\frac{D+\eta }{2}},  \label{cond}
\end{equation}%
where the band is assumed wide in the sense that $W\gg g$. In the
case opposite to one given by Eq.(\ref{cond}) the Gaussian
approximation works \cite{ourTMF1}.

The estimation (\ref{f}) does not work for the systems with continuous
broken symmetry such as $XY$ or Heisenberg model; in that case some
logarithmic corrections arise. To be specific, let us consider the case of
the $XY$ model where $\varphi =\cos \theta $ with the angle $\theta $
distributed on the interval $\left[ 0,2\pi \right) $. Then, instead of Eq.(%
\ref{h}), the following asymptotic takes place
\begin{equation}
f[q^{d/2}u_{D}\left( r\right) ]\simeq -gq^{d/2}+\frac{1}{\beta }\ln \frac{%
gq^{d/2}\beta G_{D}}{2}  \label{f1}
\end{equation}%
with a constant $G_{D}$ given by $G_{D}^{-1}=\left\langle \delta \left(
\theta _{\mathbf{0}}\right) \right\rangle _{f}$, where the average is with $%
XY$ model on discrete lattice. After minimization one has, instead of Eq.(%
\ref{energy})%
\begin{eqnarray}
\mathcal{F}_{0} &=&-g+\frac{d+2}{4d}DWq_{0},  \notag \\
q_{0} &=&\left( \frac{2d}{D}\frac{T}{W}\right) ^{\frac{2}{d+2}}\left\{ \ln %
\left[ \frac{G_{D}}{2e}\frac{g}{T}\left( \frac{2d}{D}\frac{T}{W}\right) ^{%
\frac{d}{d+2}}\right] \right\} ^{\frac{2}{d+2}}.  \label{energy1}
\end{eqnarray}%
This expression can be applied, for example, for the case of two-dimensional
$XY$ model up to the Kosterlitz-Thouless transition \cite{PP,KT} where the
correlation functions decay by power-low with the distance and $d=\eta $
grows linearly with the temperature.

Using the results obtained for the fluctuon energy one can restore the
asymptotic of DOS $N\left( E\right) $ near the band edge $E=-g$. To this end
we use the Laplace transformation connecting the partition function $Z\left(
\beta \right) =\exp \left[ -\beta \mathcal{F}\left( \beta \right) \right] $
and the DOS
\begin{equation}
Z\left( \beta \right) =\int\limits_{-g}^{\infty }N\left( E\right)
e^{-\beta E}dE,
\end{equation}
inverse transformation, and the saddle point method; the
corresponding asymptotics are connected by the so called Tauberian
theorems; this approach was used by Friedberg and Luttinger
\cite{luttinger} to obtain the Lifshitz tail for disordered
systems. To employ this in our treatment, we consider the inverse
temperature $\beta $ in Eqs. (\ref{energy}), (\ref{energy1}) as a
parameter, except the constant $G_{D}$ and, possibly, critical
exponents. Acting similarly to the derivation of Eq.(2.19) of
Ref.\cite{luttinger} one can find the Lifshitz tail in the
critical point for the discrete order parameter
\begin{equation}
N\left( E\right) \propto \exp \left[ -\left( \frac{DW}{4\epsilon }\right)
^{d/2}\ln G_{D}\right]   \label{tail}
\end{equation}
and for the $XY$ model%
\begin{equation}
\ln N\left( E\right) \propto -\left( \frac{DW}{4\epsilon }\right)
^{d/2}\left\{ \ln \left[ \frac{g}{W}\left( \frac{W}{\epsilon }\right)
^{d/2}G_{D}\right] \right\} ^{\frac{1}{d+1}},  \notag
\end{equation}%
where in both cases $\epsilon =E+g$.

Our final results mean that for the case of discrete order parameter such as
in the Ising model the main difference between the Lifshitz tails at the
critical point and above the critical region is just a replacement of the
space dimensionality $D$ by the anomalous space dimensionality $d=D-2+\eta $%
. For the case of continuous order parameter additional logarithmic
stretching of Lifshitz exponent emerge.


\begin{thebibliography}{99}
\bibitem{brinkman} W. F. Brinkman and T. M. Rice, Phys. Rev. B \textbf{2},
1324 (1970).

\bibitem{krivoglaz} M. A. Krivoglaz, Uspekhi Fiz. Nauk \textbf{111}, 617
(1973) [Engl. Transl.: Sov. Phys. Uspekhi \textbf{16}, 856 (1973)].

\bibitem{ourTMF1} M. I. Auslender and M. I. Katsnelson, Teor. Matem. Fizika
\textbf{43}, 261 (1980) [Engl. Transl.: Theor. Math. Phys. \textbf{43}, 450
(1980)].

\bibitem{ourJMMM} M. I. Auslender and M. I. Katsnelson, J. Magn. Magn.
Mater. \textbf{241}, 117 (1981).

\bibitem{ourTMF2} M. I. Auslender and M. I. Katsnelson, Teor. Matem. Fizika
\textbf{51}, 436 (1982); M. I. Auslender and M. I. Katsnelson,
Solid State Commun. \textbf{44}, 387 (1982).

\bibitem{nagaev} E. L. Nagaev, \textit{Physics of Magnetic Semiconductors}
(Mir, Moscow, 1983); E. L. Nagaev, Phys. Rep. \textbf{346}, 388 (2001).

\bibitem{dagotto} E. Dagotto, T. Hotta, and A. Moreo, Phys. Rep. \textbf{344}%
, 1 (2001).

\bibitem{Eu1} C. S. Snow, S. L. Cooper, D. P. Young, Z. Fisk, A.
Comment, and J.-P. Ansermet, Phys. Rev. B {\bf 64}, 174412 (2001).

\bibitem{Eu2} M. J. Calder\'{o}n, L. G. L. Wegener, and P. B.
Littlewood, Phys. Rev. B {\bf 70}, 092408 (2004).

\bibitem{He} A. Casey, H. Patel, J. Ny\'{e}ki, B. P. Cowan, and J.
Saunders, Phys. Rev. Lett. {\bf 90}, 115301 (2003).

\bibitem{BEC}  M. Greiner, O. Mandel, T. Esslinger, T. W. H\"{a}nsch, and I.
Bloch, Nature {\bf 415}, 39 (2003).

\bibitem{mott} N. F. Mott, \textit{Metal-Insulator Transitions} (Taylor and
Francis, London, 1974).

\bibitem{lifshitz} I. M. Lifshitz, S. A. Gredeskul, and L. A. Pastur,
\textit{Introduction to the Theory of Disordered Systems }(Wiley, New York,
1988).

\bibitem{VK} S. V. Vonsovsky and M. I. Katsnelson, \textit{Quantum Solid
State Physics }(Springer, New York, 1989).

\bibitem{feynman} R. P. Feynman, Phys. Rev. \textbf{97}, 660 (1955).

\bibitem{path1} R. P. Feynman and A. R. Hibbs, \textit{Quantum Mechanics and
Path Integrals} (McGraw Hill, New York, 1965).

\bibitem{path2} L. S. Schulman, \textit{Techniques and Applications of Path
Integration} (Wiley, New York, 1981).

\bibitem{path3} H. Kleinert, \textit{Path Integrals in Quantum Mechanics,
Statistics and Polymer Physics} (World Scientific, Singapore, 1995).

\bibitem{PP} A. Z. Patashinskii and V. L. Pokrovskii, \textit{Fluctuation
Theory of Phase Transitions} (Pergamon, New York, 1979).

\bibitem{KT} J. M. Kosterlitz and D. J. Thouless, J. Phys. C \textbf{6, }%
1181 (1973).

\bibitem{luttinger} R. Friedberg and J. M. Luttinger, Phys. Rev. B \textbf{12%
}, 4460 (1975).
\end{thebibliography}
\end{document}